\documentclass[twocolumn,showpacs,aps,prb,superscriptaddress,floatfix]{revtex4}
\usepackage{graphicx}
\usepackage{dcolumn}
\usepackage{bm}
\begin{document}
\preprint{}
%

\author{J.P. Clancy}
\affiliation{Department of Physics and Astronomy, McMaster University, Hamilton, Ontario, Canada L8S 4M1}

\author{J.P.C. Ruff}
\affiliation{Department of Physics and Astronomy, McMaster University, Hamilton, Ontario, Canada L8S 4M1}

\author{S.R. Dunsiger}
\affiliation{Department of Physics and Astronomy, McMaster University, Hamilton, Ontario, Canada L8S 4M1}

\author{Y. Zhao}
\affiliation{Department of Physics and Astronomy, McMaster University, Hamilton, Ontario, Canada L8S 4M1}

\author{H.A. Dabkowska}
\affiliation{Department of Physics and Astronomy, McMaster University, Hamilton, Ontario, Canada L8S 4M1}

\author{J.S. Gardner}
\affiliation{NIST Center for Neutron Research, NIST, Gaithersburg, Maryland 20899-8102, USA}
\affiliation{Indiana University, 2401 Milo B. Sampson Lane, Bloomington, Indiana 47408, USA}

\author{Y. Qiu}
\affiliation{NIST Center for Neutron Research, NIST, Gaithersburg, Maryland 20899-8102, USA}
\affiliation{Department of Materials Science and Engineering, University of Maryland, College Park, MD 20742, USA}

\author{J.R.D. Copley}
\affiliation{NIST Center for Neutron Research, NIST, Gaithersburg, Maryland 20899-8102, USA}

\author{T. Jenkins}
\affiliation{NIST Center for Neutron Research, NIST, Gaithersburg, Maryland 20899-8102, USA}

\author{B.D. Gaulin}
\affiliation{Department of Physics and Astronomy, McMaster University, Hamilton, Ontario, Canada L8S 4M1}
\affiliation{Canadian Institute for Advanced Research, 180 Dundas Street W., Toronto, Ontario, Canada M5G 1Z8}

\newcommand{\HTO}{Ho$_2$Ti$_2$O$_7$}
\newcommand{\DTO}{Dy$_2$Ti$_2$O$_7$}

\title{Revisiting Static and Dynamic Spin Ice Correlations in Ho$_2$Ti$_2$O$_7$}

\begin{abstract}
Elastic and inelastic neutron scattering studies have been carried out on the pyrochlore magnet {\HTO}.  Measurements in zero applied magnetic field show that the disordered spin ice ground state of {\HTO} is characterized by a pattern of rectangular diffuse elastic scattering within the [HHL] plane of reciprocal space, which closely resembles the zone boundary scattering seen in its sister compound {\DTO}.  Well-defined peaks in the zone boundary scattering develop only within the spin ice ground state below $\sim$ 2 K.  In contrast, the overall diffuse scattering pattern evolves on a much higher temperature scale of $\sim$ 17 K.  The diffuse scattering at small wavevectors below [001] is found to vanish on going to {\bf Q}=0, an explicit signature of expectations for dipolar spin ice.  Very high energy-resolution inelastic measurements reveal that the spin ice ground state below $\sim$ 2 K is also characterized by a transition from dynamic to static spin correlations on the time scale of 10$^{-9}$ seconds.  Measurements in a magnetic field applied along the [1${\bar1}$0] direction in zero-field cooled conditions show that the system can be broken up into orthogonal sets of polarized $\alpha$ chains along [1${\bar1}$0] and quasi-one-dimensional $\beta$ chains along [110].  Three dimensional correlations between $\beta$ chains are shown to be very sensitive to the precise alignment of the[1${\bar1}$0] externally applied magnetic field.

\end{abstract}

\maketitle

\section{Introduction}

{\HTO} is a member of the rare-earth titanate, cubic pyrochlore materials, which crystallize into the Fd$\bar{3}$m space group\cite{greedan}.  This family of materials has been of great interest as the magnetic rare earth sites, the A-sites in the composition A$_2$B$_2$O$_7$, reside on a network of corner-sharing tetrahedra, the archetypical structure for geometrical frustration in three dimensions\cite{diep}.  The magnetic properties of this family of materials are very diverse, as the size and anisotropy of the relevant moments, as well as the nature of the coupling between moments on different sites, varies dramatically across the rare-earth series.  Particular interest has focused on {\HTO} and {\DTO}, both of which are characterized by strongly anisotropic, Ising magnetic moments with local [111] anisotropy, and net nearest-neighbor interactions which are ferromagnetic.  The nature of the crystal field splittings associated with Ho$^{3+}$ in the {\HTO} environment is such that large magnetic moments ($\sim$ 10 $\mu_B$ per Ho) are tightly constrained to point directly into or out-of the tetrahedra on which they reside\cite{blote,mamsurova,bramwell2000}.

This combination of a hard, local $\langle 111 \rangle$ anisotropy and coupled classical spins on the pyrochlore lattice has been well studied.  Somewhat counterintuitively, antiferromagnetic near-neighbor interactions lead to a four-sublattice Neel state, which is not frustrated.  However, an effective ferromagnetic near-neighbor interaction with local $\langle 111 \rangle$ anisotropy leads to a spin ice ground state, in which the spins on each tetrahedra are constrained by ``ice rules" which require two spins pointing in and two spins pointing out on each tetrahedron\cite{harris,GBreview}.  There are six such degenerate configurations per tetrahedron, and hence an extensive number of configurations for an extended pyrochlore lattice.  As is well known, this degeneracy is the same as that which arises from proton disorder in solid water ice, and accordingly this ground state is referred to as spin ice\cite{harris}.  A unit cell of the pyrochlore lattice, containing 16 Ho$^{3+}$ ions at the A-sites, is shown in Figure 1.  The spins in Figure 1 have been chosen to obey the two-in, two-out ice rules for each tetrahedron, hence the spin configuration depicted is one of many degenerate spin ice ground states.

\begin{figure}
\includegraphics{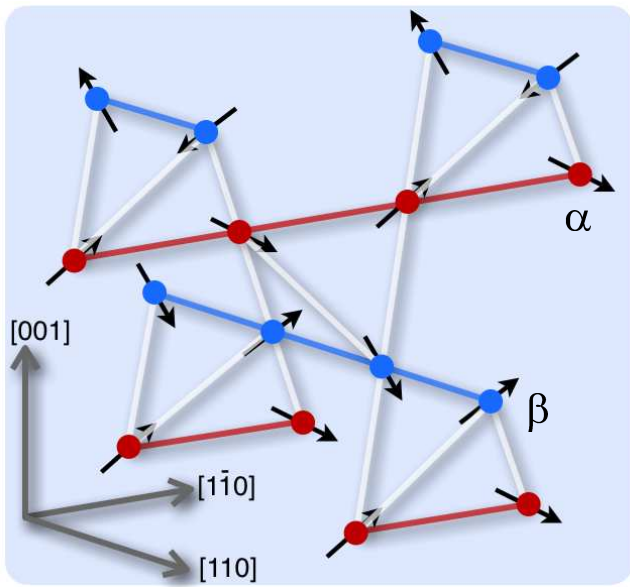}
\caption{(Color online) Magnetic moments which satisfy the ``ice rules" decorate a conventional unit cell of the pyrochlore lattice.  The structure can be thought of as comprised of two orthogonal sets of chains running along the [1$\bar{1}$0] and [110] directions.  These are shown as red (dark) spins on the $\alpha$ chains and blue (light) spins on the orthogonal $\beta$ chains.}
\end{figure}

The pyrochlore lattice can be decomposed in several fashions for convenience.  For example, it can be thought of as an alternating stacking of triangular (edge-sharing triangles) and kagome (corner-sharing triangles) layers along the [111] direction\cite{moessner111,cornelius,fennell07,higa03}.  It can also be decomposed into two orthogonal sets of chains, referred to as $\alpha$ and $\beta$ chains, which run along the orthogonal [1$\bar{1}$0] and [110] directions\cite{harris,hiroi,yoshida,ruff,fennell05}.  In this scenario, half of the spins in the system reside on the $\alpha$ chains, while half reside on the $\beta$ chains.  An illustration of this conceptual picture can be found in Figure 1, and as will become clear later on, such a decomposition can be very useful in the presence of a [1$\bar{1}$0] magnetic field which breaks the crystal symmetry along one particular [110] direction.

A useful starting point description for {\HTO} is the following Hamiltonian for spins interacting on the pyrochlore lattice, known as the standard dipolar spin ice model\cite{GBreview,ruff,denhertog,yavorskii}:

\begin{eqnarray}
{\cal H}&=&\;\; -J\sum_{\langle i,j\rangle}{\bf S}_{i}^{z_{i}}\cdot{\bf S}_{j}^{z_{j}} 
-\mu\sum_{i}{\bf S}_{i}^{z_{i}}\cdot{\bf H} \\
&+&  DR_{{\rm nn}}^{3}\sum_{ \begin{array}{c} i>j \end{array} } \frac{{\bf S}_{i}^{z_{i}}\cdot{\bf S}_{j}^{z_{j}}}{|{\bf R}_{ij}|^{3}} - \frac{3({\bf S}_{i}^{z_{i}}\cdot{\bf R}_{ij}) ({\bf S}_{j}^{z_{j}}\cdot{\bf R}_{ij})}{|{\bf R}_{ij}|^{5}} \nonumber \\ \nonumber
\end{eqnarray} 

The first term in Eq. 1 describes the relatively weak nearest-neighbor exchange interactions between Ho$^{3+}$ moments, while the second term accounts for the Zeeman energy which arises due to the interactions of Ho$^{3+}$ moments with the applied magnetic field.  The third term describes the long-range dipolar interactions, which can be of considerable importance since the size of the moments is very large.  The ${\bf S}_{j}^{z_{j}}$ notation indicates that the local Ising z-axis varies from site to site.  Good estimates have been made for both J and D in {\HTO}, yielding values of J$\sim$ -1.65 K (antiferromagnetic) and D$\sim$ 1.41 K\cite{GBreview,denhertog,bramwell2001}.  As described by Bramwell and Gingras, the antiferromagnetic near-neighbor exchange in {\HTO} is sufficiently weak that it is overcome by the near-neighbor dipolar interactions, resulting in an effective near-neighbor interaction which is ferromagnetic\cite{GBreview,denhertog,bramwell2001}.  Hence the relevant ground state for {\HTO} is expected to be the spin ice state, a prediction which has now been well verified\cite{GBreview, harris, bramwell2001}.

{\HTO} displays two low temperature anomalies in its heat capacity: a very low temperature feature which originates from nuclear hyperfine contributions, and a Schottky-like anomaly at $\sim$ 2 K, which signifies the entry to the spin ice state\cite{bramwell2001}.  The characteristics of the spin ice ground state in {\HTO}, as well as the analogous state in {\DTO}, have been of great interest for more than a decade.  Recent theoretical work has focused on the possibility of emergent cluster correlations\cite{fennell04,yavorskii}, as well as the existence of magnetic monopoles\cite{monopole} which fractionalize from the dipole moments originating from excitations out of the spin ice ground state.  Recent experimental work has investigated plateaus in the magnetization of {\HTO} and {\DTO}\cite{higa03}, and sought to identify the nature of the relevant field-induced phases and phase boundaries\cite{hiroi, ruff, moessner111, higa03, higa05, fennell05, monopole, fennell07, tabata}.  A number of experiments have also probed the unusual dynamics of spin freezing in these materials\cite{matsu2000,matsu2001,snyder,ehlers,snyder04,ehlers04}.

In this paper, we investigate several distinct themes.  First we discuss the details of S({\bf Q}) as measured with time-of-flight neutron spectroscopy at low temperatures and compare these results to previous experimental and theoretical work.  We focus specifically on topics not previously discussed: zone boundary scattering in {\HTO} and the characteristic signatures of dipolar spin ice.  We also investigate how S({\bf Q}) evolves with temperature as the spin ice state is destroyed upon warming to T$\sim$ 2 K and above.  We explore the dynamical behaviour of the Ho$^{3+}$ spins as the spin ice state is approached upon cooling, and show direct evidence for spin freezing on a 10$^{-9}$ second time scale.  Finally, we characterize the decomposition of the spin ice state in {\HTO} into polarized $\alpha$ chains coexisting with quasi-one-dimensional $\beta$ chains in the presence of a [1$\bar{1}$0] magnetic field.

\section{Experimental Details}

Several large single crystal samples of {\HTO} were grown by floating zone image furnace techniques at McMaster University.  The details of these growths were very similar to those which have previously been reported for Tb$_2$Ti$_2$O$_7$\cite{gardner98}.  The high quality single crystals of {\HTO} which resulted from these growths were $\sim$ 5 grams in mass and had approximate cylindrical dimensions of 25 mm in length and 6 mm in diameter.  A series of neutron scattering measurements, utilizing both time-of-flight and backscattering techniques, were performed on two of these {\HTO} single crystals at the National Institute of Standards and Technology Center for Neutron Research.  In both of these experiments the samples were aligned such that the [HHL] plane in reciprocal space was coincident with the horizontal scattering plane.

Time-of-flight neutron scattering measurements were performed using the Disk Chopper Spectrometer (DCS).  DCS uses a series of seven choppers to create pulses of monochromatic neutrons, whose energy transfers on scattering are determined from their arrival times at the instrument's 913 detectors, located at scattering angles ranging from -30 to 140 degrees\cite{DCS}.  Measurements were performed using both 5 {\AA} and 9 {\AA} incident neutrons.  These measurements were used to map out the pattern of diffuse scattering in the [HHL] plane as a function of temperature and magnetic field, as well as to search for inelastic scattering with approximate energy resolutions of $\sim$ 0.1 meV (5 {\AA} measurements) and 0.02 meV (9 {\AA} measurements).

Much higher energy-resolution measurements were also carried out using the High Flux Backscattering Spectrometer (HFBS). In HFBS neutrons are backscattered from a spherically focusing monochromator, interact with the sample, and are backscattered a second time by a spherically focusing analyzer before reaching a bank of 16 detectors ranging from {\bf Q} = 0.1 to 2.0 {\AA}$^{-1}$ [Ref. 31].  A Phase Space Transformation (PST) chopper is used to increase the neutron flux of the instrument by Doppler shifting the incident neutron wavelength distribution towards the appropriate backscattered wavelength. A high speed Doppler drive is used to oscillate the monochromator, significantly expanding the dynamic range of the instrument.  In this experiment, inelastic measurements were performed with a dynamic range of $\pm$ 17 $\mu$eV and an approximate energy resolution of 0.92 $\mu$eV.  This is roughly a factor of 20 times greater energy resolution than the 9 {\AA} measurements collected with DCS.

\section{Diffuse Magnetic Scattering in Zero Magnetic Field}

Figures 2 a) and b) show S({\bf Q}) for {\HTO} at T$\sim$ 0.2 K measured in the [HHL] plane.  Figure 2 a) was collected using DCS and 5 {\AA} incident neutrons and 2 b) with DCS and 9 {\AA} incident neutrons, allowing smaller $\vert$Q$\vert$ scattering to be accessed.  The time-of-flight neutron scattering technique simultaneously measures both the {\bf Q} and energy dependence of the scattering.  However, the diffuse scattering in both Figures 2 a) and b) is found to be resolution-limited in energy, and hence is elastic in nature.  The diffuse scattering can be seen to peak up into well-defined rectangular regions, the most prominent of which are centered on the (0,0,1) and (0,0,3) positions in reciprocal space.  The main region in the field of view which is devoid of diffuse scattering is centered on (0,0,2) and can be seen to have an approximately hexagonal shape.  The overall pattern closely resembles the zone boundary scattering which has been reported for {\DTO}\cite{fennell04}, suggesting that such scattering may be characteristic of spin ice as manifested in both of these materials.  In {\DTO}, this zone boundary scattering has been attributed to either emergent hexagonal clusters\cite{fennell04}, or to expectations from a generalized dipolar spin ice model which includes further-neighbour exchange interactions\cite{yavorskii}.  The measured S({\bf Q}) for {\HTO} in Figure 2 differs in detail from both earlier reported neutron diffraction measurements on {\HTO}\cite{bramwell2001}, and from the expectations of the near-neighbor spin ice model and the simple dipolar spin ice model (Eq. 1)\cite{GBreview,bramwell2001,yavorskii}.

\begin{figure}
\includegraphics{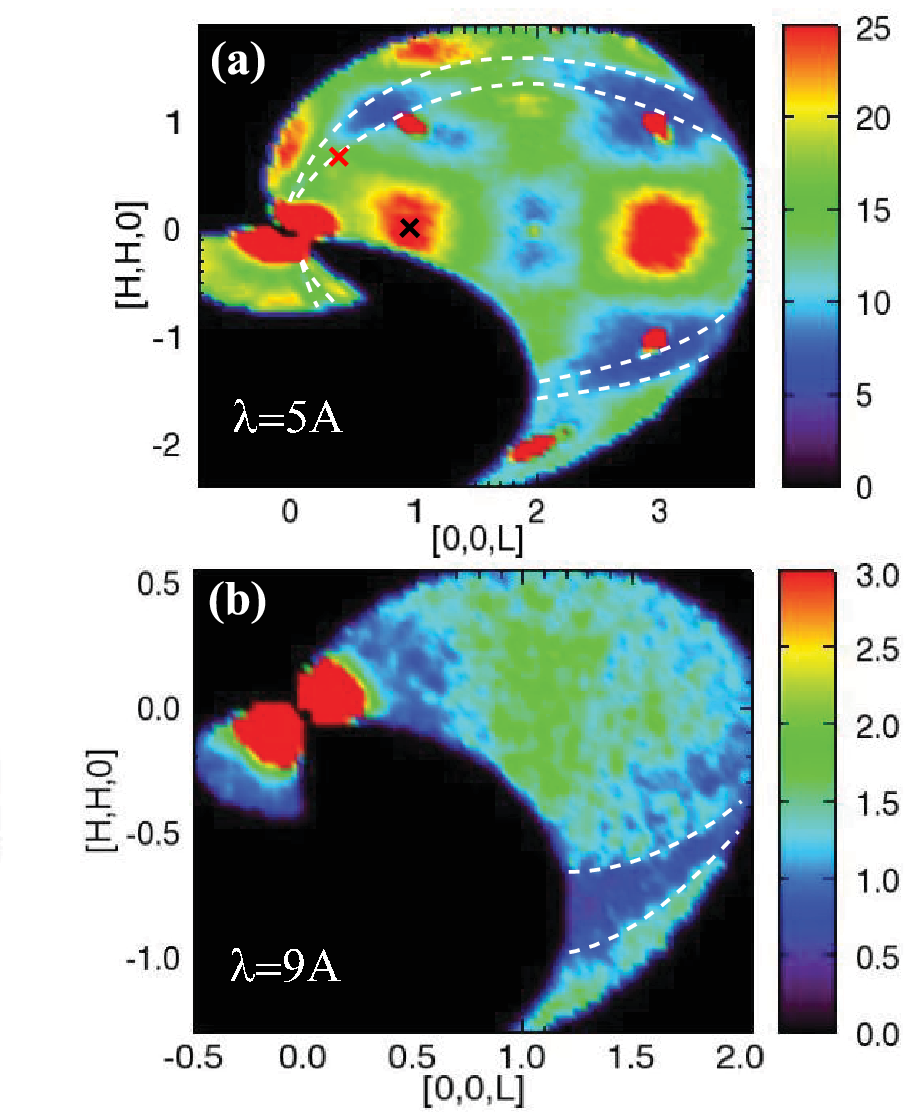}
\caption{(Color online) Elastic scattering maps of S({\bf Q}) are shown for {\HTO} at T=0.2 K.  Measurements in a) employed 5 {\AA} incident neutrons and elastic scattering corersponds to $\Delta$E$\le$0.1 meV.  b) employed 9 {\AA} incident neutrons and elastic scattering corresponds to $\Delta$E$\le$0.02 meV.  The ``x"s indicate the {\bf Q} positions at which high energy-resolution backscattering measurements were performed.  Dashed white lines indicate the dark angle caused by the sample environment.}
\end{figure}

One of the most striking differences between calculations for the near-neighbor spin ice model (with net near-neighbor ferromagnetic exchange and hard $\langle 111 \rangle$ anisotropy) and the simple dipolar spin ice model (which includes the full Hamiltonian shown in Eq. 1) is a pronounced depression of the diffuse magnetic scattering near the origin of reciprocal space, at {\bf Q}=0\cite{GBreview,bramwell2001}.  This region of negligible scattering intensity at small {\bf Q} arises due to the presence of dipolar interactions, which preclude the formation of a state with non-zero magnetization in zero applied field.  This feature is unmistakably visible in our new measurements on {\HTO}, shown in Figure 2 b), which clearly reveal that the diffuse scattering drops off to background for L$\le$0.6 for {\bf Q}= [HHL].  The direct observation of this low {\bf Q} depression in scattering intensity provides compelling evidence for the dipolar nature of the correlations in the ground state of {\HTO}.  This feature of S({\bf Q}) can only be observed in the present set of measurements because they employ long wavelength 9 {\AA} incident neutrons.  

Zone boundary \textit{inelastic} scattering has been observed in the spinel ZnCr$_2$O$_4$\cite{lee} where it has been interpreted in terms of strongly correlated spins in approximately independent hexagonal clusters which reside within the kagome planes of the pyrochlore structure.  Similar scattering features have also been identified in the spinels CdFe$_2$O$_4$\cite{kamazawa} and CdCr$_2$O$_4$\cite{chung}.  Theoretical work on a generalized dipolar spin ice model has shown that elastic zone boundary scattering can also be accounted for by tuning exchange interactions, not included in Eq. 1, beyond near-neighbors\cite{yavorskii}.  The importance of further-neighbor exchange interactions in spin ice has also been demonstrated by studies in which up to third nearest-neighbour exchange couplings have been used to explain the critical temperatures of observed field-induced transitions in {\DTO}\cite{ruff, moessner111, higa05}. In any case, it seems clear that the nature of the zone boundary scattering is characteristic of a host of such highly frustrated pyrochlore magnets.

S({\bf Q}) measurements using DCS with 5 {\AA} incident neutrons, similar to those shown in Figure 2 a), were performed as a function of temperature up to T=20 K in order to investigate the evolution of spatial correlations across the spin ice transition near 2 K.  As mentioned earlier, the entrance to the spin ice ground state is identified by a Schottky-like anomaly in the heat capacity, so no conventional phase transition is expected.  An additional surprisingly high temperature scale has also been identified - at least in {\DTO}.  AC susceptibility measurements on {\DTO} reveal a frequency-dependent anomaly in the 15 K regime\cite{snyder,matsu2001}.  A corresponding anomaly has also been identified in {\HTO}, but only in the presence of a DC field\cite{ehlers}.  This relaxation process is believed to be a thermally activated single-ion process, with an activation energy which is governed by the gap between the ground state and the first excited crystal field state\cite{ehlers, snyder04}.  The observed spin fluctuation rate is also typical for an individual spin flip process.  It is the presence of these anomalies in the AC susceptibility that determined the temperature scale on which we chose to explore the evolution of S({\bf Q}) in {\HTO}.

Representative S({\bf Q}) measurements for {\HTO} in the [HHL] plane at T=2 K (close to the spin ice Schottky-anomaly), T=4 K (well above the spin ice anomaly), and T=20 K (above the high temperature scale identified in the AC susceptibility measurements) are shown in Figures 3 a), b), and c), respectively.  These maps can be compared to the T=0.2 K data provided in Figure 2 a) to show how the zero field S({\bf Q}) in {\HTO} develops with temperature, from an order of magnitude below the specific heat anomaly at T $\sim$ 2 K to an order of magnitude above it.  As can be seen from these maps, the evolution of the measured S({\bf Q}) is somewhat subtle.  This is not so surprising as the disordered spin ice ground state at T=0.2 K already displays exceptionally short correlation lengths (with spins correlated over a single tetrahedron), and hence warming up into the conventional paramagnetic phase cannot substantially shorten the overall spin correlations.  The integrated intensity of line scans taken through the peak in the diffuse scattering at (0,0,3) is shown in Figure 3 d).  This shows that the integrated diffuse scattering falls off gradually with slight upwards curvature on a relatively high temperature scale of $\sim$ 17 K - very close to the 15 K anomaly identified in the AC susceptibility measurements.  This sets the scale on which S({\bf Q}) can evolve, since beyond $\sim$ 17 K there is very little structure to S({\bf Q}) within our field of view.  It should be noted that neutron spin echo measurements have detected evidence of diffuse magnetic scattering in {\HTO} up to temperatures as high 800 K, although any indications of spatial correlations appear to vanish above $\sim$55 K\cite{ehlers04}.

\begin{figure}
\includegraphics{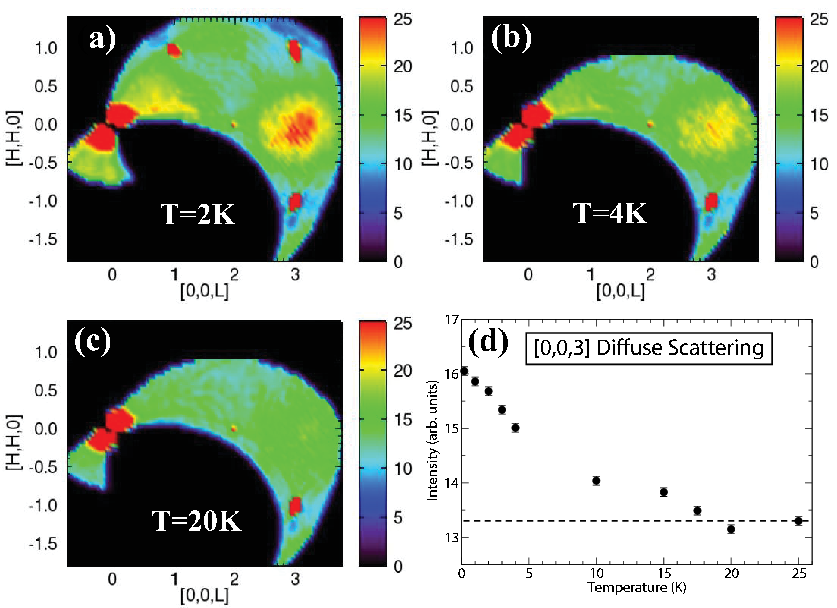}
\caption{(Color online) S({\bf Q}) for {\HTO} in the [HHL] plane is shown at zero magnetic field for a) T=2 K, b) T=4 K, and c) T=20 K.  The temperature evolution of the diffuse scattering at (0,0,3) is shown in d). Diffuse scattering intensities have been obtained by integrating over cuts taken along the [HH3] and [00L] directions.  The dashed line provided in d) is intended as a guide-to-the-eye, illustrating the plateau in diffuse scattering intensity above $\sim$ 17 K.  The error bars in d) and in all subsequent figures represent the standard deviation in the measurement.}
\end{figure}

A collection of representative line scans taken through the zone boundary scattering are provided in Figure 4.  These scans have been taken along the [-2H+4/9, -2H+4/9, H+16/9] direction, allowing them to cut through the zone boundary scattering at its sharpest points, near (2/3, 2/3, 5/3) and (-2/3, -2/3, 7/3).  The orientation of these cuts with respect to the overall pattern of diffuse scattering is displayed in the inset to Figure 4 a), where the direction of the line scans can be identified by the dashed magenta line from (1,1,3/2) to (-1,-1,5/2).  As a guide-to-the-eye, the hexagonal shape of the zone boundaries has also been highlighted by a series of dashed white lines.  The line scans provided in the main panel of Figure 4 a) illustrate how the zone boundary scattering evolves with temperature in zero field, with data sets collected at T=0.2 K, 2 K, and 20 K. These cuts clearly demonstrate that well-defined peaks in the diffuse zone boundary scattering can only be identified below the spin ice anomaly at T$\sim$ 2 K.  Figure 4 b) shows similar scans collected in the spin ice ground state of {\HTO} at T=0.2 K, with a magnetic field applied along the [1$\bar{1}$0] direction.  As will be discussed in the following section, the application of a [1$\bar{1}$0] field decouples the system into orthogonal $\alpha$ and $\beta$ chains, destroying the spin ice ground state at sufficiently high field strengths.

\begin{figure}
\includegraphics{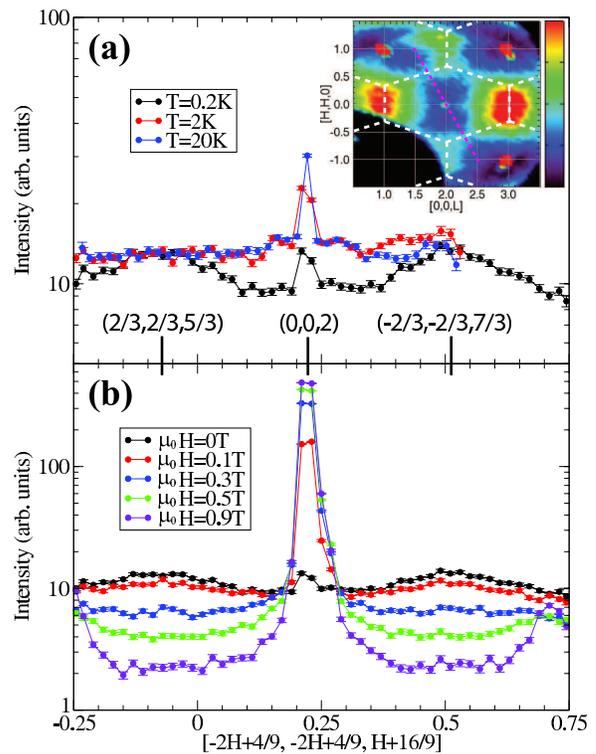}
\caption{(Color online) Representative cuts taken through the sharpest points of the zone boundary scattering, at (2/3, 2/3, 5/3) and (-2/3, -2/3, 7/3), showing a) the temperature evolution of the scattering at $\mu_0$H=0 T, and b) the magnetic field evolution of the scattering at base temperature (T $\sim$ 0.2 K). All cuts are taken along the [-2H+4/9, -2H+4/9, H+16/9] direction. The orientation of these cuts with respect to the [HHL] plane is illustrated by the dashed magenta line in the inset to panel a).  The zone boundaries have been outlined in white as a guide-to-the-eye.}
\end{figure}

An interesting subject which has recently attracted considerable attention is the potential presence of ``pinch point" scattering in {\HTO}.  Pinch point features have been theoretically predicted to arise in frustrated magnets as a consequence of enforcing the local ice rules\cite{henley,fennell07,huse}.  These features, which are relatively sharp in specific directions, could be expected to broaden significantly upon passing through the spin ice anomaly.  Sharp pinch point features in S({\bf Q}) have already been identified in {\HTO} in the presence of a [111] magnetic field\cite{fennell07}.  Such an applied magnetic field gives rise to a magnetization plateau which signifies the stability of the kagome ice phase of {\HTO}\cite{moessner111,higa03}.  As previously mentioned, the pyrochlore lattice can be decomposed into a series of alternating triangular and kagome planes which are stacked normal to the [111] direction.  The kagome ice phase is characterised by spins which are locally constrained on the kagome planes, and spins which are polarized by the applied field on the triangular planes.

The pinch point scattering predicted by the dipolar spin ice model is expected to arise at Brillouin zone centers, such as (0,0,2).  However, as the map of S({\bf Q}) provided in Figure 2 a) clearly indicates, the (0,0,2) position is situated at the center of a region which is almost entirely devoid of diffuse scattering.  Thus the present set of measurements reveal no indications of pinch point correlations in {\HTO} at zero field, a result which is fully consistent with the previous work of Fennell \textit{et al}\cite{fennell07}.  Unfortunately, due to the presence of a large field-dependent Bragg peak at (0,0,2) and the emergence of rod-like scattering along [00L], we are unable to offer any definitive comment on the potential for pinch point features in an applied [1$\bar{1}$0] field.  Any evidence for the relatively subtle pinch point correlations is likely to be obscured by these much stronger field-dependent scattering features.

\begin{figure}
\includegraphics{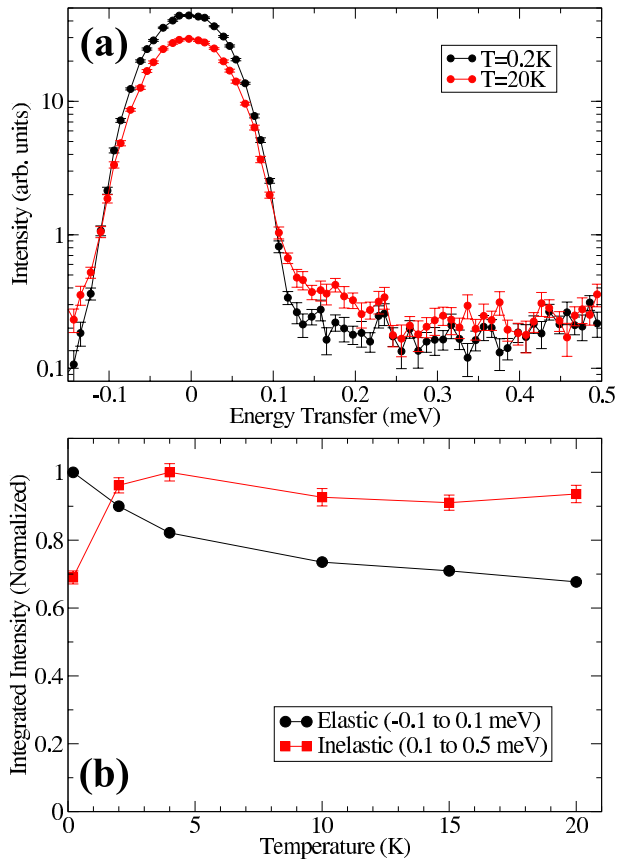}
\caption{(Color online) a) Representative energy cuts taken through zero field data sets at T=0.2 K and 20 K. Each cut has been binned over H=[-0.4,0.4] and L=[2.6,3.4] in order to capture the diffuse scattering at (0,0,3). b) The temperature evolution of the total elastic and inelastic scattering intensity, obtained by integrating over energy scans of the form shown in panel a).}
\end{figure}

It is also interesting to investigate how signatures of the entrance into the spin ice ground state are manifested in the spin dynamics of {\HTO}.  As discussed earlier, below 2 K the energy-dependence of the diffuse magnetic scattering in {\HTO} appears to be entirely elastic on the energy resolution scale which is accessible with DCS (0.1 meV with 5 {\AA} incident neutrons; 0.02 meV with 9 {\AA} incident neutrons).  Interestingly, this is very different from the diffuse magnetic scattering characteristic of the spin liquid ground state in Tb$_2$Ti$_2$O$_7$ in zero field\cite{gard99, gard03, rule}, which is inelastic on the energy scale of $\sim$ 0.3 meV\cite{rule}.  As shown in Figure 5 a), the energy width of the diffuse scattering can be seen to broaden slightly above 2K, providing an indication that spins are becoming more dynamic outside of the spin ice state.  This is confirmed by the data in Figure 5 b), which shows that the gradual decrease in the intensity of the elastic diffuse scattering at higher temperatures is accompanied by a simultaneous increase in the intensity of the inelastic diffuse scattering.  This shows that the static spins of the spin ice state undergo slow fluctuations as the temperature of the system rises above 2K.  Improving on this energy resolution requires either the use of backscattering techniques to measure S({\bf Q},$\hbar\omega$) directly, or neutron spin echo (NSE) measurements of S({\bf Q},t) which have already been carried out on {\HTO}.  NSE measurements on {\HTO} show a continuous evolution of the time-correlations for spins within the time window to which NSE is sensitive, over a temperature scale from 0.3 K to 150 K\cite{ehlers06, ehlers08}.

We have performed backscattering measurements near two wavevectors in the [HHL] plane; at (0,0,1), which coincides with a peak in the pattern of diffuse scattering, and at (2/3, 2/3, 1/3), which lies near one of the narrowest regions of zone boundary scattering.  These two positions are indicated by the pair of ``x"s provided in Figure 2 a).  Characteristic backscattering energy scans for {\HTO} at T=1.4 K and T=5 K are shown in Figure 6. As previously observed for the DCS data in Figure 5, these representative scans clearly indicate a broadening of the energy width above the Schottky anomaly at 2 K.  In order to extract quantitative values for the energy width of the diffuse scattering these data sets were fit to Lorentzian lineshapes which were convoluted by a Gaussian resolution function.  The resulting FWHM of the scattering, which is characteristic of an inverse lifetime $\tau$, is plotted as a function of temperature in Figure 7.  We observe that the diffuse scattering only approaches our resolution limit on entering the spin ice ground state below $\sim$ 2 K.  The paramagnetic state above T=2 K is characterized by a finite spin relaxation rate which is roughly temperature independent from $\sim$ 3 to 30 K and increases with temperature above 30 K. Consequently, the distinguishing characteristics of the disordered spin ice ground state below the Schottky anomaly at 2 K are that the structure of the zone boundary scattering within S({\bf Q}) is well developed, and the structure is static on a time scale of 10$^{-9}$ seconds.  Neither of these statements is true for {\HTO} at temperatures of T=2 K or above in zero applied field.

\begin{figure}
\includegraphics{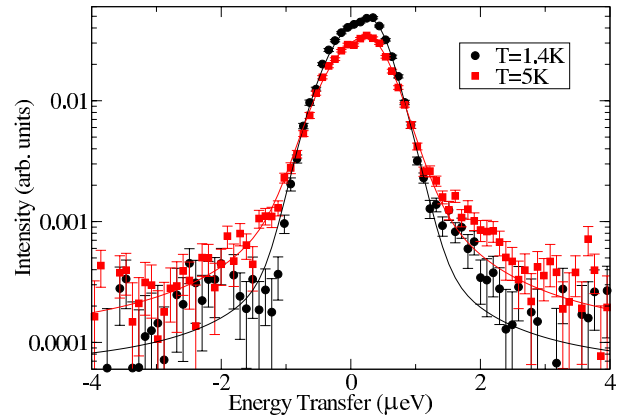}
\caption{(Color online) Representative backscattering energy scans taken at {\bf Q}=(0,0,1) in {\HTO}.  The scans displayed here were collected at T=1.4 K (within the spin ice state) and T=5 K (well above the spin ice state).}
\end{figure}

\begin{figure}
\includegraphics{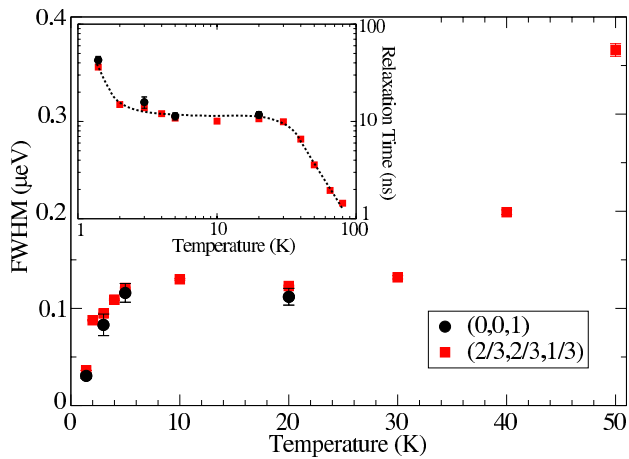}
\caption{(Color online) Parameters extracted from fits to backscattering data obtained at the (0,0,1) and (2/3, 2/3, 1/3) positions in reciprocal space. Shown in the main panel is the temperature evolution of the FWHM of the quasi-elastic signal. Provided in the inset is the temperature dependence of the relaxation time, $\tau$, which is inversely proportional to the width in energy. The spin ice state in {\HTO} is characterised by spins which are static on the 10$^{-9}$ second scale.}
\end{figure}

In the inset to Figure 7 we see the temperature evolution of the characteristic spin relaxation time $\tau$ plotted on a log-log scale.  Over the larger temperature window provided in this inset we can see three distinct relaxation regimes: (i) the high temperature Arrhenius-like relaxation above 30K, (ii) the temperature independent so-called quantum relaxation from 3K to 30K, and (iii) the diverging relaxation times indicative of spin-freezing below 2K.  Plotted in this fashion, our data above 2 K fall into excellent agreement with previously reported NSE measurements on {\HTO}\cite{ehlers08}.  However, what is unique to the present {\HTO} backscattering measurements is our observation of the dramatic increase in relaxation times which occurs below the specific heat anomaly at 2 K.  This behavior is fully consistent with the observed temperature dependence of the characteristic spin relaxation time in {\DTO} as measured by AC susceptibility\cite{snyder03}, with only a slight shift in the relevant temperature scales distinguishing the two materials. 

Our backscattering measurements can also provide information about how the integrated intensity of the diffuse scattering evolves with temperature.  In particular, the backscattering data collected at the (0,0,1) position are a useful complement to the DCS measurements shown in Figure 3 d) which describe the temperature dependence of the scattering at (0,0,3).  Both (0,0,1) and (0,0,3) correspond to peaks in the overall pattern of diffuse scattering, and the consistency between the two sets of measurements is very good.  In both cases the integrated intensity of the diffuse scattering appears to drop off on a much higher temperature scale than the relaxation time, somewhere in the vicinity of 15 to 20 K.

\section{Scattering under H$\parallel$[1$\bar{1}$0]  Zero-Field-Cooled Conditions}

It has been appreciated for some time that the application of a magnetic field along [110] and equivalent directions should decouple the spin ice ground state into polarized $\alpha$ chains (which point along the field) and perpendicular $\beta$ chains\cite{harris,hiroi,yoshida,ruff,fennell05}, as illustrated in Figure 1.  This decoupling has been demonstrated directly in {\DTO} by neutron scattering experiments\cite{fennell05}.  Related measurements have also been performed on {\HTO}\cite{fennell05}, although to date maps of the low temperature diffuse scattering have only been reported for {\DTO}.

In Figure 8 the measured S({\bf Q}) for {\HTO} in the [HHL] plane is shown at base temperature (T=0.2 K) for H$\parallel$[1$\bar{1}$0].  The data sets provided in these four panels correspond to applied magnetic field strengths of a) 0.2 T, b) 0.4 T, c) 0.9 T, and d) 2.5 T.  When combined with the zero field S({\bf Q}) data displayed in Figure 2, these maps illustrate the evolution of the diffuse scattering from broad, rectangular, zone boundary scattering at zero field to narrow, rod-like scattering extending along [00L] at finite fields.  This rod-like scattering can be interpreted as arising from sheets of quasi-one-dimensional scattering from decoupled $\beta$ chains intersecting with the [HHL] scattering plane.  Coinciding with the development of this rod-like scattering is the growth of the scattering intensity at the (0,0,2) Bragg peak, which is indicative of the polarization of the $\alpha$ chains.  Although it is possible to have ordered magnetic states with polarized $\alpha$ chains and negligible (0,0,2) peak intensity, in the case where there is no net moment within the $\beta$ chain sublattice, the intensity of the (0,0,2) peak can be taken as a direct measure of the $\alpha$ chain polarization.  For these measurements, and all subsequent measurements reported below, the magnetic field was applied at low temperatures following a zero-field cooled protocol.

\begin{figure}
\includegraphics{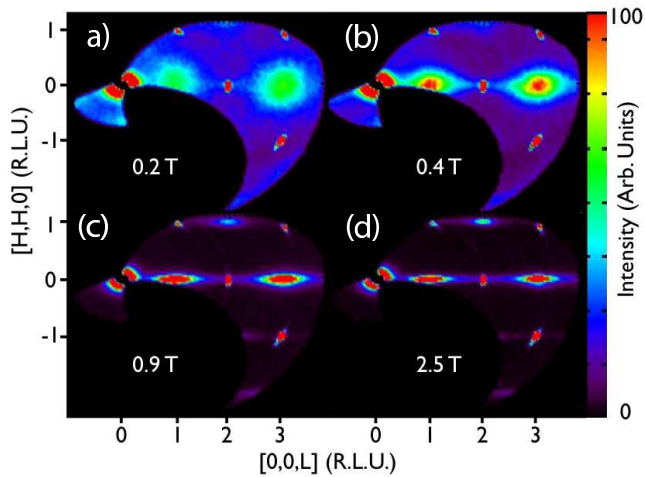}
\caption{(Color online) S({\bf Q}) for {\HTO} in the [HHL] plane is shown for T=0.2 K and H$\parallel$[1$\bar{1}$0].  The maps displayed above correspond to field strengths of a) 0.2 T, b) 0.4 T, c) 0.9 T and d) 2.5 T.  In each case the magnetic field was applied following a zero-field cooled protocol. }
\end{figure}

Cuts through the diffuse scattering shown in Figure 8 were taken along the [00L] direction, parallel to the rod of scattering, and along the [HH3] direction, perpendicular to the rod, as shown in Figures 9 a) and b).  These cuts were then fit to an anisotropic Lorentzian using an Ornstein-Zernike form, for the purpose of extracting correlation lengths parallel and perpendicular to the rods.  The resulting correlation lengths in real space are shown as a function of field in Figure 9 d).  These fits show that the diffuse scattering is fully three-dimensional in zero field, as expected, with identical correlation lengths of $\sim$ 3.6 {\AA}, or one near-neighbor separation, in all directions.  The scattering immediately becomes anisotropic upon the application of a [1$\bar{1}$0] magnetic field, with both sets of correlation lengths growing as a function of increasing field strength.  As can be seen in Figure 9 d), the transition to a quasi-one-dimensional structure is complete by $\sim$ $\mu_0$H=1 T, at which point the correlations along the $\beta$ chains are resolution-limited ($>$ 100 {\AA}) and those between $\beta$ chains are saturated at 10 {\AA}, or roughly two interchain distances.  The integrated intensity of the (0,0,2) Bragg scattering was also monitored as a function of [1$\bar{1}$0] magnetic field, as shown in Figure 9 c).  As noted earlier, the (0,0,2) intensity can be interpreted as a measure of the polarization of spins on the $\alpha$ chains, and it is observed to grow sharply and approach saturation by 0.9 T. This corresponds very closely with the full development of the quasi-one-dimensional correlations of spins residing on the $\beta$ chains, a result which agrees very well with previous neutron results\cite{fennell05}.

\begin{figure}
\includegraphics{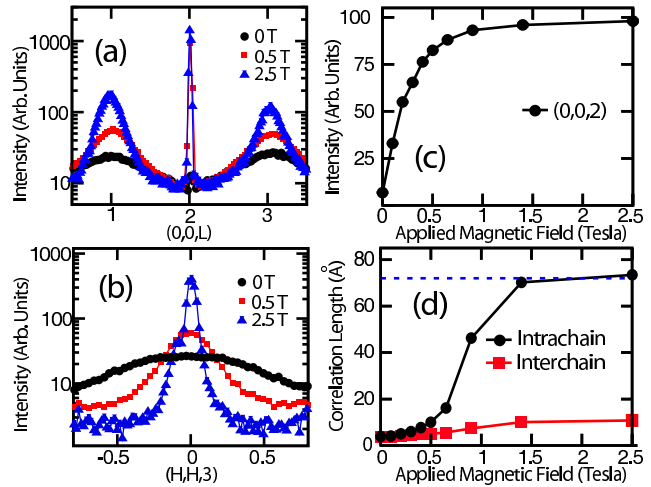}
\caption{(Color online) Representative cuts of the data shown in Fig. 8 taken along a) [00L] and b) [HH3] are shown at T=0.2 K for applied magnetic fields of $\mu_0$H=0, 0.5, and 2.5 T.  The field dependence of the (0,0,2) Bragg peak intensity is shown in c). The correlation lengths of the diffuse scattering along [00L] and [HH3] are shown in d), where the dashed line indicates the resolution limit of the instrument.  The quasi-one-dimensional correlations of the $\beta$ chains are established by a $\sim$ 1 T magnetic field $\parallel$[1$\bar{1}$0].}
\end{figure}

We now return to the zone boundary scattering shown in the line scans of Figure 4, and specifically how these spin ice correlations evolve with the application of a [1$\bar{1}$0] magnetic field.  The field dependence of the zone boundary scattering is shown in Figure 4 b) for T=0.2 K and representative field strengths of 0, 0.1, 0.3, 0.5, and 0.9 T, respectively.  These scans capture the peaks from the narrowest regions of the diffuse scattering, near (2/3,2/3,5/3) and (-2/3,-2/3,7/3), as well as the field-dependent Bragg scattering observed at (0,0,2).  We find that the intensity of the zone boundary scattering begins to drop with the application of any finite [1$\bar{1}$0] magnetic field, with evidence of distinct peaks vanishing by $\sim$ 0.5 T.  This coincides with the point at which the (0,0,2) Bragg scattering, representing the polarization of spins on the $\alpha$ chains, is almost fully developed.  The complete elimination of the diffuse scattering along the cuts shown in Figure 4 b) requires a field of 0.9 T, at which point the intensity of the (0,0,2) peak is completely saturated.

We also note that the scattering intensity of the (0,0,2) Bragg reflection is weak, but unmistakably non-zero, even in the absence of an applied magnetic field.  This (0,0,2) scattering can clearly be observed in the zero-field measurements of S({\bf Q}) provided in Figures 2, 3, and 4.  This weak Bragg feature is structurally forbidden within the Fd$\bar{3}$m space group appropriate to the pyrochlores, and appears to be temperature independent up to at least T=20 K.  Earlier neutron diffraction measurements on pyrochlores such as Tb$_2$Ti$_2$O$_7$\cite{rule,hyperfine} and Er$_2$Ti$_2$O$_7$\cite{ruffeto} have identified similar scattering at the (0,0,2) position, but in general there has been a tendency to associate this feature with harmonic Bragg ($\lambda$/n) contamination.  Since the present time-of-flight measurements (like those of Refs. 39 and 44) do not employ Bragg reflection from a monochromator cystal, we can be assured that such scattering is not due to higher order contamination, and likely indicates a weak departure from the perfect pyrochlore structure in {\HTO} and perhaps in all real pyrochlore materials\cite{rule}.

A series of S({\bf Q}) measurements were also performed for {\HTO} with the applied magnetic field offset from [1$\bar{1}$0] by a rotation of 7.6 $\pm$ 3 degrees about the [00L] direction.  These measurements should be contrasted with the data discussed above for which the magnetic field is nominally applied precisely along [1$\bar{1}$0], but in practice is aligned to within an accuracy of $\pm$ 1 degree.  The identification of the precise misalignment of the magnetic field comes about by virtue of the relative positions at which several Bragg peaks were observed on the three detector banks of DCS. In this case, the appearance of the (0,0,2) Bragg reflection on the middle detector bank, combined with allowed Fd$\bar{3}$m Bragg peaks such as (1,1,3) and (-1,-1,3) appearing on the upper and lower detector banks, is consistent with a field misaligned from [1$\bar{1}$0] by 7.6 $\pm$ 3 degrees about [00L].  Note that the same rotation which changes the orientation of the applied field will also affect the scattering plane used in the experiment.  If the applied field direction is offset from [1$\bar{1}$0] by some angle $\theta$ about [00L], then the new scattering plane can be expressed as:
\begin{eqnarray}
[H{\left( cos\theta + sin\theta \right)}, H{\left( cos\theta - sin\theta \right)}, L] 
\end{eqnarray} 
Thus, for a field offset of $\theta$ $\sim$ 7.6 degrees, the scattering plane should be roughly coincident with the [1.13H, 0.86H, L] plane in reciprocal space.

A comparison of the measured S({\bf Q}) for {\HTO} in the presence of aligned and misaligned applied magnetic fields is provided in Figure 10.  Both of the data sets depicted here were collected using the same crystal of {\HTO}, and both sets of measurements were taken at T=0.2 K and $\mu_0$H=0.5 T.  Figure 10 a) shows the nominally aligned data, where we can see quasi-one-dimensional rods of scattering along [00L], but clear short-range correlations along the [HH0] $\beta$ chain direction, consistent with the data shown in Figure 4.  In addition, one can clearly observe the strong (0,0,2) Bragg scattering which arises due to the polarization of $\alpha$ chains.  Figure 10 b) shows the same data set with the applied field canted by 7.6 $\pm$ 3 degrees with respect to the [00L] direction.  Cuts along the [00L] direction for the data sets shown in Figures 10 a) and b) are overlaid in c).  It is evident from b) and c) that the diffuse scattering along [00L] is far more one-dimensional in the presence of the canted magnetic field, with much less structure visible in both the S({\bf Q}) maps and the cuts taken through the data.  Interestingly, while the properties of the [00L] diffuse scattering are significantly altered by field misalignment, the nature of the Bragg scattering at (0,0,2) remains essentially unchanged.

\begin{figure}
\includegraphics{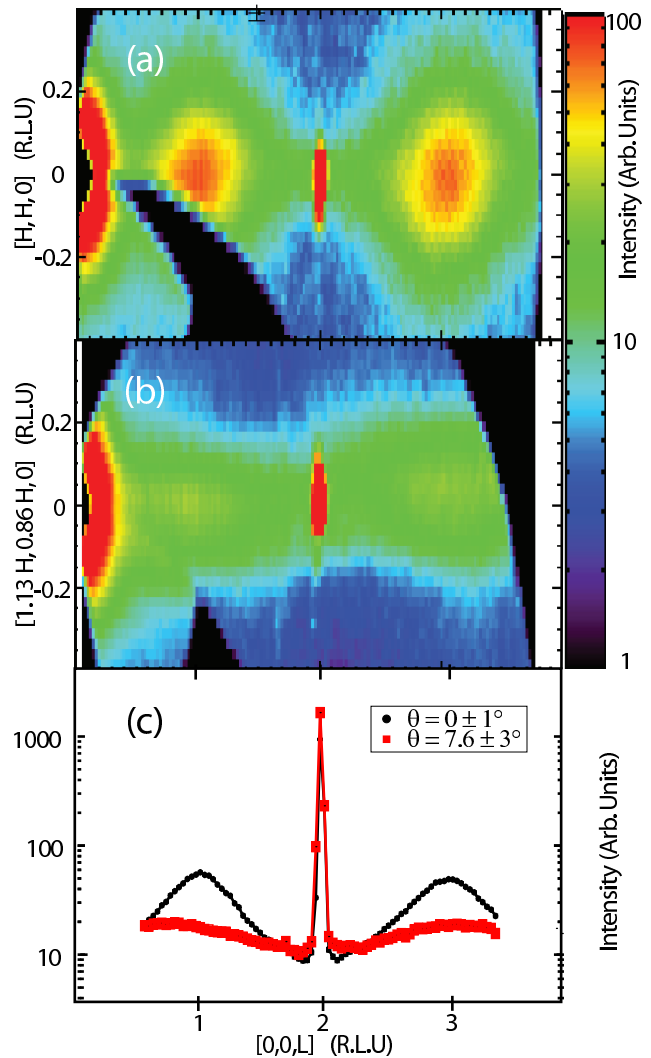}
\caption{(Color online) A comparison is made between diffuse scattering with a) H applied $\parallel$ [1$\bar{1}$0], and b) H canted with respect to [1$\bar{1}$0] by 7.6 $\pm$ 3 degrees.  The canted field implies a field component along the $\beta$ chains which frustrates their short range ordering and results in a more-one-dimensional structure.  The vertical axes in a) and b) are a function of $\theta$, the angle between [1$\bar{1}$0] and the direction of the applied field, which can be written as [H(cos$\theta$+sin$\theta$), H(cos$\theta$-sin$\theta$), 0].  The cuts provided in c) have been taken along the [00L] direction, through the diffuse scattering shown in a) and b).}
\end{figure}

The effect of the canted [1$\bar{1}$0] field can be understood in terms of its decomposition into a large aligned [1$\bar{1}$0] field along the $\alpha$ chains, and a smaller field perpendicular to the $\alpha$ chains due to misalignment.  In this case, the perpendicular component of the field is aligned along [110], parallel to the $\beta$ chains.  The weak, primarily dipolar, interactions between $\beta$ chains are expected to be antiferromagnetic in nature, and this is consistent with the pattern of diffuse scattering observed in Figure 8.  The application of a small uniform magnetic field along the direction of the $\beta$ chains would frustrate these antiferromagnetic correlations, resulting in the enhanced one-dimensionality which we observe in the presence of the canted field.  We also expect that small offsets to the field direction should have little measureable effect on the polarization of the $\alpha$ chains, since the component of the field along [1$\bar{1}$0] will be only slightly diminished by misalignment. Once again, this prediction appears to be fully consistent with the data shown in Figure 10.

One implication of this result is that such offsets to the field orientation, even in the case of nominal ``perfect" alignment, may be sufficient to frustrate the three dimensional correlations between $\beta$ chains. In turn, this frustration may cause the spin ice system to remain in a quasi-one-dimensional short-range ordered state, rather than arriving at its true equilibrium state\cite{ruff}, characterized by long-range ordered $\beta$ chains.  The significance of small deviations from perfect alignment has also been demonstrated theoretically by Monte Carlo studies on dipolar spin ice\cite{melko04}, where misalignments of only one degree have been shown to suppress antiferromagnetic interchain order in modest [1$\bar{1}$0] applied fields.

\section{Conclusions}

In conclusion, we have carried out neutron scattering measurements on the spin ice ground state of single crystal {\HTO} as a function of temperature and applied magnetic field.  In zero magnetic field, S({\bf Q}) at T=0.2 K is characterized by diffuse, elastic scattering which closely resembles the zone boundary elastic scattering seen in {\HTO}'s sister spin ice material, {\DTO}.  We have explicitly shown that the diffuse scattering falls to zero as $\vert$Q$\vert$ goes to zero, consistent with theoretical predictions from the dipolar spin ice model.  We have also shown that the diffuse zone boundary scattering develops a well-defined structure only within the spin ice state for T$<$ 2 K.  Furthermore, it has been shown that the integrated intensity of the strongest diffuse features, the rectangles centered at (0,0,1) and (0,0,3), evolves on a surprisingly high characteristic temperature scale of $\sim$ 17 K, similar to that of the 15 K anomaly observed in {\DTO}.  Very high energy-resolution inelastic neutron scattering measurements have demonstrated that the spin ice ground state of {\HTO} is static on a time scale of 10$^{-9}$ seconds.

We have also measured the evolution of S({\bf Q}) at T=0.2 K for {\HTO} in the presence of a [1$\bar{1}$0] magnetic field.  Application of the field decomposes the system into spins residing on polarized $\alpha$ chains and those residing on weakly correlated $\beta$ chains, leading to the formation of a quasi-one-dimensional magnetic substructure.  We have shown that the peaks in the diffuse scattering at the zone boundaries vanish for magnetic field strengths which substantially polarize the $\alpha$ chains.  Finally, we have demonstrated that the nature of the three-dimensional correlations between quasi-one-dimensional $\beta$ chains is very sensitive to the precise alignment of the externally applied [1$\bar{1}$0] magnetic field.  We hope that this work will help to guide and inform future studies of the spin ice state in the rare earth pyrochlores.

\begin{acknowledgments}

The authors would like to acknowledge helpful discussions with M.J.P. Gingras, as well as useful contributions from A.B. Kallin and E.M. Mazurek.  This work was supported by NSERC of Canada and utilized facilities supported in part by the National Science Foundation under Agreement No. DMR-0454672.  The DAVE software package was used for elements of the data reduction and analysis described in this paper\cite{dave}.

\end{acknowledgments}

%
%
%
%
%
%
%
%
%
%


\begin{thebibliography}{}
\bibitem{greedan}
for a review, see:  J.E. Greedan, \textit{Journal of Alloys and Compounds}, Vol. 408-412, pp. 444 (2006).
\bibitem{diep}
\textbf{Frustrated Spin Systems}, edited by H.T. Diep (World Scientific Publishing Co. Pte. Ltd., Singapore, 2004). 
\bibitem{blote}
H.W.J. Blote, R.F. Wielinga, and W.J. Huiskamp,  Physica (Amsterdam) {\bf 43}, 549 (1969).
\bibitem{mamsurova}
L.G. Mamsurova, K. K. Pukhov, N. G. Trusevich, and L. G. Shcherbakova.  Sov. Phys. Solid State {\bf 27}, 1214 (1985).
\bibitem{bramwell2000}
S.T. Bramwell, M.N. Field, M.J. Harris, I.P. Parkin, J. Phys. Condens. Matter {\bf 12}, 483 (2000).
\bibitem{harris}
M.J. Harris, S.T. Bramwell, D.F. McMorrow, T. Zeiske, K.W. Godfrey, Phys. Rev. Lett. {\bf 79}, 2554 (1997).
\bibitem{GBreview}
S.T. Bramwell and M.J.P. Gingras, Science {\bf 294}, 1495 (2001). 
\bibitem{moessner111}
R. Moessner, S.L. Sondhi, Phys. Rev. B {\bf 68}, 064411 (2003).
\bibitem{cornelius}
A.L. Cornelius and J.S. Gardner, Phys. Rev. B {\bf 64}, 060406 (2001).
\bibitem{fennell07}
T. Fennell, S.T. Bramwell, D.F. McMorrow, P. Manuel, A.R. Wildes, Nature Physics {\bf 3}, 566 (2007).
\bibitem{higa03}
R. Higashinaka, H. Fukazawa and Y. Maeno, Phys. Rev. B {\bf 68}, 014415 (2003).
\bibitem{hiroi}
Z. Hiroi, K. Matsuhira and M. Ogata.  J. Phys. Soc. Jpn. {\bf 72}, 3045 (2003).
\bibitem{yoshida}
S.-I. Yoshida, K. Nemoto and K. Wada.  J. Phys. Soc. Jpn. {\bf 73}, 1619 (2004).
\bibitem{ruff}
J.P.C. Ruff, R.G. Melko and M.J.P. Gingras, Phys. Rev. Lett. {\bf 95}, 097202 (2005).
\bibitem{fennell05}
T. Fennell, O.A. Petrenko, B. Fak, J.S. Gardner, S.T. Bramwell, and B. Ouladdiaf, Phys. Rev. B {\bf 72}, 224411 (2005).
\bibitem{denhertog}
B.C. den Hertog and M.J.P. Gingras, Phys. Rev. Lett. {\bf 84}, 3430 (2000).
\bibitem{yavorskii}
T. Yavors'kii, T. Fennell, M.J.P. Gingras and S.T. Bramwell, Phys. Rev. Lett. {\bf 101}, 037204 (2008).
\bibitem{bramwell2001}
S.T. Bramwell, M.J. Harris, B.C. den Hertog, M.J.P. Gingras, J.S. Gardner, D.F. McMorrow, A.R. Wildes, A.L. Cornelius, J.D.M. Champion, R.G. Melko, and T. Fennell.  Phys. Rev. Lett. {\bf 87}, 047205 (2001).
\bibitem{fennell04}
T. Fennell, O.A. Petrenko, B. Fak, S.T. Bramwell, M. Enjalran, T. Yavors'kii, M.J.P. Gingras, R.G. Melko and G. Balakrishnan.  Phys. Rev. B {\bf 70}, 13440 (2004).
\bibitem{monopole}
C. Castelnovo, R. Moessner, and S.L. Sondhi, Nature {\bf 451}, 42 (2008).
\bibitem{higa05}
R. Higashinaka and Y. Maeno, Phys. Rev. Lett. {\bf 95}, 237208 (2005).
\bibitem{tabata}
Y. Tabata, H. Kadowaki, K. Matsuhira, Z. Hiroi, N. Aso, E. Ressouche, and B. Fak, Phys. Rev. Lett. {\bf 97}, 257205 (2006).
\bibitem{matsu2000}
K. Matsuhira, Y. Hinatsu, K. Tenya, T. Sakakibara, J. Phys. Condens. Matter {\bf 12}, L649 (2000). 
\bibitem{matsu2001}
K. Matsuhira, Y. Hinatsu, T. Sakakibara, J. Phys. Condens. Matter {\bf 13}, L737 (2001). 
\bibitem{snyder}
J. Snyder, J. S. Slusky, R. J. Cava, P. Schiffer, Nature {\bf 413}, 48 (2001).
\bibitem{ehlers}
G. Ehlers, A.L. Cornelius, M. Orendac, M. Kajnakova, T. Fennell, S.T. Bramwell and J.S. Gardner.   J. Phys: Condens. Matter, {\bf 15}, L9 (2003).
\bibitem{snyder04}
J. Snyder, B.G. Ueland, A. Mizel, J.S. Slusky, H. Karunadasa, R.J. Cava, and P. Schiffer, Phys. Rev. B {\bf 70}, 184431 (2004).
\bibitem{ehlers04}
G. Ehlers, A.L. Cornelius, T. Fennell, M. Koza, S.T. Bramwell, and J.S. Gardner, J. Phys: Condens. Matter {\bf 16}, S635 (2004).
\bibitem{gardner98}
J.S. Gardner, B.D. Gaulin and D. McK Paul, J. Crys. Growth {\bf 191}, 740 (1998).
\bibitem{DCS}
J.R.D. Copley and J.C. Cook, Chem. Phys. {\bf 292}, 477 (2003). 
\bibitem{HFBS}
A. Meyer, R.M. Dimeo, P.M. Gehring, and D.A. Neumann, Rev. Sci. Instrum. {\bf 74}, 2759 (2003).
\bibitem{lee}
S.-H. Lee, C. Broholm, W. Ratcliff, G. Gasparovic, Q. Huang, T.H. Kim, and S.-W. Cheong, Nature {\bf 418}, 856 (2002).
\bibitem{kamazawa}
K. Kamazawa, S. Park, S.-H. Lee, T.J. Sato and Y. Tsunoda.  Phys. Rev. B {\bf 70}, 024418 (2004).
\bibitem{chung}
J.-H. Chung, M. Matsuda, S.-H. Lee, K. Kakurai, H. Ueda, T.J. Sato,H. Takagi, K.-P. Hong and S. Park.  Phys. Rev. Lett. {\bf 95}, 247204 (2005).
\bibitem{henley}
C. L. Henley, Phys. Rev. B {\bf 71}, 014424 (2005).
\bibitem{huse}
D.A. Huse, W. Krauth, R. Moessner, and S.L. Sondhi, Phys. Rev. Lett. {\bf 91}, 167004 (2003).
\bibitem{gard99}
J.S. Gardner, S.R. Dunsiger, B.D. Gaulin, M.J.P. Gingras, J.E. Greedan, R.F. Kiefl, M.D. Lumsden, W.A. MacFarlane, N.P. Raju, J.E. Sonier, I. Swainson, and Z. Tun.  Phys. Rev. Lett. {\bf 82} 1012 (1999).
\bibitem{gard03}
J.S. Gardner, A. Keren, G. Ehlers, C. Stock, Eva Segal, J.M. Roper, B. Fak, M.B. Stone, P.R. Hammar, D.H. Reich, and B.D. Gaulin.  Phys. Rev. B {\bf 68}, 180401 (2003).
\bibitem{rule}
K.C. Rule, J.P.C. Ruff, B.D. Gaulin, S.R. Dunsiger, J.S. Gardner, J.P. Clancy, M.J. Lewis, H.A. Dabkowska, I. Mirebeau, P. Manuel, Y. Qiu and J.R.D. Copley.  Phys. Rev. Lett. {\bf 96}, 177201 (2006).
\bibitem{ehlers06}
G. Ehlers, J.S. Gardner, C.H. Booth, M. Daniel, K.C. Kam, A.K. Cheetham, D. Antonio, H.E. Brooks, A.L. Cornelius, S.T. Bramwell, J. Lago, W. Haussler, and N. Rosov, Phys. Rev. B {\bf 73}, 174429 (2006).
\bibitem{ehlers08}
G. Ehlers, J.S. Gardner, Y. Qiu, P. Fouquet, C.R. Wiebe, L. Balicas, and H.D. Zhou.  Phys. Rev B {\bf 77}, 052404 (2008).
\bibitem{snyder03}
J. Snyder, B.G. Ueland, J.S. Slusky, H. Karunadasa, R.J. Cava, A. Mizel, and P. Schiffer, Phys. Rev. Lett. {\bf 91}, 107201 (2003).
\bibitem{hyperfine}
J.S. Gardner and A. Hoser, Hyperfine Interactions {\bf 133}, 269 (2001).
\bibitem{ruffeto}
J.P.C. Ruff, J.P. Clancy, A. Bourque, M.A. White, M. Ramazanoglu, J.S. Gardner, Y. Qiu, J.R.D. Copley, H.A. Dabkowska, B.D. Gaulin, arXiv.org:0808.1082 (2008).
\bibitem{melko04}
R.G. Melko and M.J.P. Gingras, J. Phys: Condens. Matter {\bf 16}, R1277 (2004).
\bibitem{dave}
http://www.ncnr.nist.gov/dave.

\end{thebibliography}
\end{document}